# High-dimensional Assisted Generative Model for Color Image Restoration

Kai Hong, Chunhua Wu, Cailian Yang, Minghui Zhang, Yancheng Lu, Yuhao Wang, *Senior Member, IEEE*, Qiegen Liu, *Senior Member, IEEE*

*Abstract*—This work presents an unsupervised deep learning scheme that exploiting high-dimensional assisted score-based generative model for color image restoration tasks. Considering that the sample number and internal dimension in score-based generative model have key influence on estimating the gradients of data distribution, two different high-dimensional ways are proposed: The channel-copy transformation increases the sample number and the pixel-scale transformation decreases feasible space dimension. Subsequently, a set of high-dimensional tensors represented by these transformations are used to train the network through denoising score matching. Then, sampling is performed by annealing Langevin dynamics and alternative data-consistency update. Furthermore, to alleviate the difficulty of learning high-dimensional representation, a progressive strategy is proposed to leverage the performance. The proposed unsupervised learning and iterative restoration algorithm, which involves a pre-trained generative network to obtain prior, has transparent and clear interpretation compared to other data-driven approaches. Experimental results on demosaicking and inpainting conveyed the remarkable performance and diversity of our proposed method.

*Index Terms*—Color image restoration, generative model, high-dimensional tensor, progressive strategy, unsupervised learning.

## I. INTRODUCTION

Image restoration (IR) aims to precisely reconstruct the missing or damaged parts of images from incomplete data, which is usually known to be an ill-posed problem. To resolve the issue, regularization-based techniques have been widely used by regularizing the solution spaces [1]-[6]. For an effective regularizer, it is important to find the appropriate prior knowledge of natural images [1]-[4].

Mathematically, the forward formulation of IR can be expressed as:

$$y = Mx + e \qquad (1)$$

where $x \in \mathbb{R}^N$ is the unknown image to be estimated, $y \in \mathbb{R}^N$ represents the degraded observation, and $e$ is additive noise. $M \in \mathbb{R}^{N \times N}$ denotes the degradation matrix relating to a degraded imaging system. Note that for different settings of $M$, different tasks can be presented. In this work, the primary focus is color image demosaicking and inpainting.

In the past few decades, the related works of IR in the field of image processing have thrived. Taking the image inpainting for instance, the traditional methods include partial differential equations-based methods (PDE), sparsity-based methods, and patch-matching methods, etc. For example, after Bertalmio *et al.* [7] attempted to propagate information from the outside of the inpainting region along isophotes to fill in the holes, a great amount of PDE-based inpainting techniques achieve good results for non-textured images with local thin missing regions [8]-[12]. Nevertheless, for larger missing areas, the structure in the image cannot propagate to the missing areas, which may produce blurry effects. Dabov *et al.* [13] proposed the well-known block-matching and 3D filter methods for image denoising based on sparse representation in the transform domain and a specially developed collaborative Wiener filtering. However, when using the sparsity of the image in the transform domain to reconstruct a large-area damaged image [14]-[19], the sparsity-based repair algorithm will destroy the characteristics of the image in the transform domain. In short, because of the inherent limitations of traditional methods, the diversity of images cannot be described properly and the versatility is poor.

Compared to traditional methods, many deep learning methods have been proposed to achieve efficient IR. Deep learning can generate novel content by training on large-scale datasets [20]-[21], and the visual effect is more realistic. Deep learning method can be mainly categorized as follows: Supervised learning and unsupervised learning. So far, most CNN-based deep learning approaches [22]-[24] use supervised learning. On the one hand, existing CNN-based solutions have outperformed other methods with a large margin for several simple tasks such as image denoising and super-resolution reconstruction [25]-[27]. Meinhardt *et al.* [27] proposed a set of DCNN models are pre-trained for image denoising task and are integrated into the optimization-based IR framework for different IR tasks. On the other hand, recent studies reveal that one can plug CNN-based denoisers into model-based optimization methods for solving more complicated IR tasks [28]-[30], which also promotes the widespread use of CNNs. For example, Zhang *et al.* [29] proposed a denoising convolutional neural network (DnCNN) for image denoising and general IR tasks. Furthermore, CNN denoisers can also serve as a kind of plug-and-play prior. By incorporating with unrolled inference, any restoration tasks can be tackled by sequentially applying the CNN denoisers [30]. These methods not only promote the application of CNN in low-level vision but also present many solutions to exploit CNN denoisers for other IR tasks. However, one major challenge of supervised learning is the requirement of abundant reference images

This work was supported by National Natural Science Foundation of China (61871206, 61601450).

K. Hong, H. Wu, C. Yang, M. Zhang, Y. Lu, Y. Wang, and Q. Liu are with the Department of Electronic Information Engineering, Nanchang University, Nanchang 330031, China. ({hongkai, wuchunhua, yangcailian}@email.ncu.edu.cn, ylu5@ualberta.ca, {zhangminghui, wangyuhao, liuqiegen}@ncu.edu.cn).

with labels for network training. Regretfully, finding suitable reference images becomes an obstacle for users.

To circumvent the above restriction, a variety of generative models in the manner of unsupervised learning are presented, such as autoencoder and generative adversarial networks (GAN) [31]. In general, these approaches use a coarse-to-fine network, such as the contextual attention module to guarantee the performance for inpainting [32]-[36]. Autoencoder is a type of unsupervised generative model for learning efficient data encoding. Similarly, Kingma *et al.* [37] proposed the autoencoder-based variants methods (VAE) and it has been widely used in embedding and decoding Digits [38]-[40], Faces [41]-[42], and more recent CIFAR images [38]. Furthermore, Goodfellow *et al.* [31] first proposed the GAN as a "game" for solving IR tasks. Chen *et al.* [32] proposed a GAN-based semantic image progressive inpainting method, a pyramid strategy from a low-resolution image to higher one is performed for repairing the image. However, GAN is known to suffer from unstable training and lack of consistency, which often obstruct a quantitative comparison of models. Recently, the method of score-based generative model is extensively studied. Especially, Song *et al.* [43] proposed a score-based generative model (NCSN), which represents probability distributions through the score. The score function is learned via denoising score matching (DSM) [44]-[46] and without adversarial optimization. It perturbs the data using multiple noise scales so that the scoring network captures both coarse and fine-grained image features. They combined the information from all noise scales by sampling from each noise-perturbed distribution sequentially with Langevin dynamics, and it retained decent results.

Based on the above observations, the main purpose of this study is to apply a score-based generative model assisted by high-dimensional signals. Considering that the representation ability is influenced by sample number and internal dimension, the score-based matching network estimates the corresponding data density in two high-dimensional versions of channel-copy transformation and pixel-scale transformation, respectively. Additionally, it is sampled by annealing Langevin dynamics in high-dimension tensor. The above process formulates a high-dimensional assisted generative model (HGM), in which the high-dimensionality of correlative channel features and the flexibility of the generative model achieve promising performance comprehensively. For example, at the bottom line of Fig. 1(d), we can observe the sparse of the difference between the result of HGM and the original image. Particularly, the high-dimensional structures can represent distinct but complementary information that exists at various scales. Thus, using information in lower dimension as additional guide to higher dimension for solving IR task is applicable, namely progressive HGM (HGM-p). As shown in the top line of Fig. 1(e), it achieves encouraging performance. The main reason is that the generative model is inevitably restricted by the training distribution, and the generated missing information cannot be guaranteed to strictly meet the transformation relationship.

In summary, the main contributions of this work are as follows:
- **High-dimensional strategy:** To resolve the influence of the sample number and internal dimension on estimating the gradients of data distribution, high-dimensional tensors are constructed through the correlation between channels to obtain multi-channel prior information.
- **Progressive iterative scheme:** To enhance network stability using HGM on different tasks, a progressive iterative strategy (HGM-p) is developed. According to the aggregation principle, the progressive iterative scheme avoids falling into local minima and makes the iterative process to be more robust.

It is worth noting that, this work is a spiritually significant development of our previously presented work of grayscale image restoration in [56]. At first, in [56], multi-channel and multi-model-based autoencoding prior that learned from color images is employed to grayscale image restoration. By contrast, by executing simple manipulations on color images itself in this work, we exploit the high-dimensional assisted generative model for color image restoration. More importantly, under rigorous mathematical derivation from the representation boundary of prior information, the underlying benefits of these manipulations are clearly explained.

The rest of this paper is presented as follows. Section II briefly describes some relevant works on score-based generative model. In section III, we elaborate on the formulation of the proposed method and its main characteristics. Section IV presents the IR performance of the present model, including the image demosaicking and inpainting comparisons with the state-of-the-arts. Discussions and future works are given in Section V and VI, respectively.

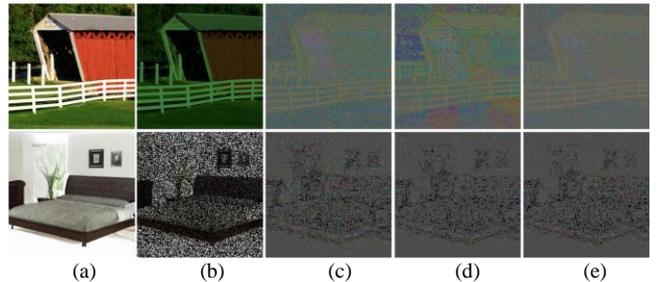

(a) (b) (c) (d) (e)

**Fig. 1.** Visualization of the difference between the results of HGM and original image in two typical IR reconstructions (Top: demosaicking, Bottom: inpainting). From left to right: (a) Ground truth, (b) Observation, (c) NCSN, (d) HGM-copy, (e) HGM-copy-p.

## II. PRELIMINARY

### A. Score Matching for Score Estimation

Generative models have shown outstanding performance in modeling data distributions which utilize known input invariances. Those models might form good prior knowledge when launching parameter learning. Most existing generative models optimize log-likelihood or do adversarial training to describe the data distribution as identical as possible. Score matching is a method which estimates the score function of the data distribution by optimizing a score network $s_\theta(x)$ parameterized with $\theta$. It is originally designed for learning non-normalized statistical models based on samples from an unknown data distribution.

Later, Song *et al.* [51] presented NCSN that trained a score network $s_\theta(x)$ directly to estimate the gradient of data density $\nabla_x \log p(x)$, instead of data density $p(x)$. More specifically, utilizing the expected squared error between $s_\theta(x)$ and $\nabla_x \log p_{data}(x)$, the function of $\theta$ can be expressed as:

$$E_{x \sim p_{data}(x)}[\| s_\theta(x) - \nabla_x \log p_{data}(x) \|_2^2] \qquad (2)$$

where $E_{x \sim p_{data}(x)}[\cdot]$ is approximated as the sample mean over $x$. In practice, using data samples with denoising score matching (DSM) can efficiently calculate expectation over $p_{data}(x)$ in Eq. (2). The idea is to perturb $x$ in accordance with a pre-specified noise distribution $q_\sigma(\tilde{x}|x)$ (i.e., $E_{q_\sigma(\tilde{x}|x)p(x)}[\|s_\theta(\tilde{x}) - \nabla_{\tilde{x}} \log q_\sigma(\tilde{x}|x)\|_2^2]$). Empirically speaking, DSM is suited as it takes less time and naturally fits the task of estimating scores with noise-perturbed data distributions. Then, estimating the score of the distribution $q_\sigma(\tilde{x}) = \int q_\sigma(\tilde{x}|x) p_{data}(x) dx$ in the perturbed version. Assuming the noise distribution is chosen to be $q_\sigma(\tilde{x}|x) = N(\tilde{x}|x, \sigma^2 I)$, therefore $\nabla_{\tilde{x}} \log q_\sigma(\tilde{x}|x) = -(\tilde{x}-x)/\sigma^2$. For a given $\sigma$, Eq. (2) can be written as:

$$E_{x \sim p_{data}(x), \tilde{x} \sim N(\tilde{x}|x, \sigma^2 I)}[\|s_\theta(\tilde{x}) + (\tilde{x}-x)/\sigma^2\|_2^2] \quad (3)$$

Then, Eq. (3) is combined for all $\sigma \in \{\sigma_i\}_{i=1}^L$ to get one unified objective:

$$L(\theta) = \frac{1}{2L} \sum_{i=1}^{L} E_{p_{data}(x)} E_{p_{\sigma_i}(\tilde{x}|x)}[\|\sigma_i s_\theta(\tilde{x}, \sigma_i) + (\tilde{x}-x/\sigma_i)\|_2^2] \quad (4)$$

As a conical combination of DSM objectives, one can imagine that Eq. (4) achieves the minimum value if and only if $s_{\theta^*}(x, \sigma_i) = \nabla_x \log p_{\sigma_i}(x)$ for all $i \in \{1, 2, \cdots, L\}$. Indeed, perturbing the data by multiple levels noise to train a single conditional score network, and estimating the score of the corresponding levels greatly improve the performance of the score-based generative model.

*B. Langevin Dynamics*

Langevin dynamics can produce samples from a probability density $p(x)$ only using the score function $\nabla_x \log p(x)$. Starting from an initial point $x_0$, Langevin dynamics can iteratively refine it in a noisy gradient ascent fashion so that the value of log-density $\log p(x)$ will increase:

$$x_{t+1} \leftarrow x_t + \frac{\alpha}{2} \nabla_x \log p(x_t) + \sqrt{\alpha} z_t \quad (5)$$

where $\alpha > 0$ is a step size, $z_t$ is a standard Gaussian white noise, and $T$ is the total number of iterations. It can be proved from Eq. (5), when $\alpha$ is sufficiently small and $T$ is sufficiently large, $x_t$ will be an exact sample from $p(x)$ under some regularity conditions. Furthermore, suppose we have a neural network $s_\theta(x, \sigma_i)$ parameterized by $\theta$, and it has been trained such that $s_\theta(x, \sigma_i) \approx \nabla_x \log p(x)$:

$$x_{t+1} \leftarrow x_t + \frac{\alpha}{2} s_\theta(x_t, \sigma_i) + \sqrt{\alpha} z_t \quad (6)$$

We can approximately generate samples from $p(x)$ using Langevin dynamics by replacing $\nabla_x \log p(x_t)$ with $s_\theta(x_t)$ in Eq. (6).

## III. PROPOSED HGM MODEL

In the previous section, the score matching for score estimation and Langevin dynamics are discussed, where the estimation is calculated from data density to the gradient of data density. In this section, we concentrate on effectively improving the performance of the native DSM on IR.

*A. Motivation*

Although the original NCSN has achieved promising results, there are still two major deficiencies: Low data density regions and the manifold hypothesis. Firstly, this model is often assumed that the data distribution is supported on a low dimensional manifold, such that matching method will fail to provide a consistent score estimator. Secondly, the lack of training data in low data density regions, e.g., far from the manifold, hinders the matching accuracy and slows down the mixing of Langevin dynamics sampling. Besides, NCSN not effectively uses the signal attributes, such as local similarity, and three-dimensional structural information. However, from the perspective of machine learning, learning high-dimensional information is conducive to generating representations. One recent work in this direction was reported in [47], where Quan *et al.* estimated the target gradient by using a high-dimensional tensor as the input to NCSN for MRI reconstruction.

Given the above two aspects, a high-dimensional assisted generative learning framework is putting forward to enhance the network stability that favors avoiding falling into local optima. More rigorously, the main idea depends on the following theorem.

**Theorem 1** [48]. Let $F$ be a class of $\mathbb{R}^d$-valued functions, all of which are $M/2$-Lipchitz, bounded coordinate wise by $R > 0$, containing arbitrarily good approximations of $\nabla \log p_\sigma$ on the ball of radius $R$. Let $\sigma < \sigma_{\max}$ and support we have $n$ i.i.d samples from $p_\sigma$, $x_1, \cdots, x_n$. Let

$$\hat{s} \in \arg\min_{s \in F} \frac{1}{n} \sum_{i=1}^{n} \|s(x_i) - \nabla \log p_\sigma(x_i)\|_2^2 \quad (7)$$

Then with probability at least $1 - 4\delta - Cne^{-R^2/m_\sigma}$ on the randomness due to the sample,

$$E_{p_\sigma}[\|\hat{s}(x) - \nabla \log p_\sigma(x)\|_2^2] \leq C(MR+B)^2 (\log^3 n \cdot \mathfrak{R}_n^2(F) + \beta_n d) \quad (8)$$

where $C$, $M$ and $B$ are universal constant. Both $\mathfrak{R}_n^2(F)$ and $\beta_n = (\log(1/\delta) + \log\log n)/n$ contain the factor of $1/n$. The right side of Eq. (8) is mainly related to the sample number $n$ and internal dimension $d$.

More specifically, there exist the following tendencies between the representation error $E_{p_\sigma}(n,d)$ of Eq. (8) and $n$, $d$, respectively:

**Lemma 1.** For any class $F$ of real-valued functions with image contained in the unit interval $[0,1]$, and let $n$ be the number of the data sample, we have:

$$E_{p_\sigma}(n_2, d) < E_{p_\sigma}(n_1, d) \quad if: n_2 > n_1 \quad (9)$$

**Lemma 2.** Let $F$ be a class of real-valued functions with image contained in the unit interval, we denote $d$ is the dimension and $d_1 > d_2$, we have:

$$E_{p_\sigma}(n, d_1) > E_{p_\sigma}(n, d_2) \quad if: d_1 > d_2 \quad (10)$$

From **Theorem 1**, the representation error is inversely proportional to sample number $n$ and is proportional to space dimension $d$. More specifically, **Lemma 1** states that under the condition of multivariate Gaussian, the sample number and the representation error are negatively correlated. From **Lemma 2**, it should be emphasized that the representation error is completely intrinsic to the geometry of the data manifold and the dimension of the feature space does not appear. Thus, it can be concluded that even with arbitrarily high-dimension in pixel-level intensity space, if the feasible space has small dimension $d$, the problem of low-dimensional manifolds and low data density regions in generating density priors will be alleviated.

For the convenience of exhibiting **Theorem 1**, the tendency relationships between $n$, $d$ and the representation error $E_{p_\sigma}(n,d)$ are visualized in Fig. 2 and Fig. 3, respectively. Our idea is motivated by the manifold hypothesis and the theoretical analysis mentioned above. In this work, we present two substantially different ways, HGM-copy and HGM-pool, to improve the effectiveness and accuracy of the score matching. More precisely, HGM-copy stacks the channel by copying transformation to increase the sample number $n$. A simple channel-copy operation will not change the distribution of the data, but it is effective in enlarging the sample. Meanwhile, HGM-pool is based on the down-sampling operator to reduce the feasible space dimension $d$ of the data, which can effectively transfer some of the hard work of score estimation to the easier, lower-dimensional regime, as well as boosting the performance of generative model. The detailed description of these two ways and implementation of the algorithms will be presented in the following context.

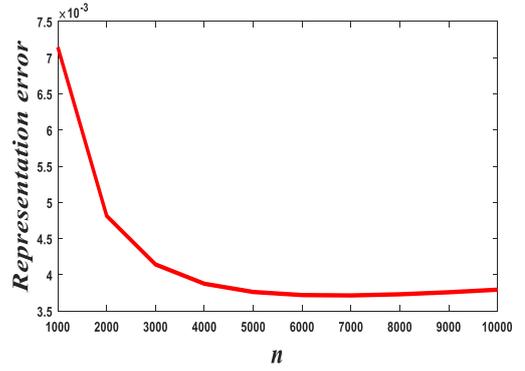

**Fig. 2.** Visualization of the representation error and the relationship between sample number $n$.

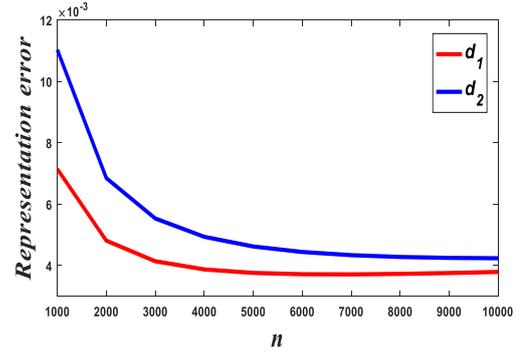

**Fig. 3.** Visualization of the representation error with regarding to different space dimensions $d$. There exists $d_1 > d_2$.

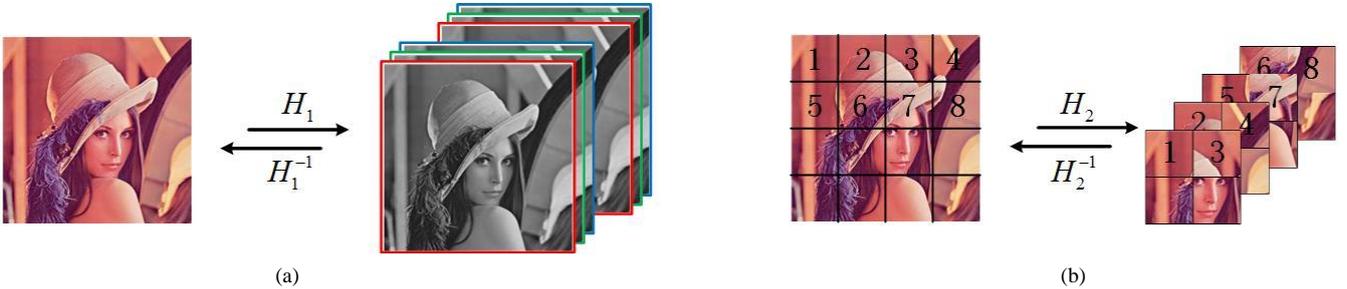

**Fig. 4.** Invertible transformations illustration of the high-dimensional data. (a) The 2-D channel-copy transformation, (b) The pixel-scale transformation that extracts the average pooling of the data distribution according to a checkerboard pattern.

### B. HGM: Prior in High-Dimensional Space

As an intuitive consideration of the above two aspects in **Lemma 1** and **2**, we propose two kinds of high-dimensional reconstruction methods to address the issue of NCSN: HGM-copy and HGM-pool.

On the one hand, to take advantage of high-dimensional tensor as an advanced prior, HGM-copy handles data by concatenating them as multi-channel objects. The components of high-dimensional tensor are set as the network input via channel-copy technology. As shown in Fig. 4(a), HGM-copy maps the data to a high-dimensional space through a channel-copy operation, rather than conducts on the original low-dimensional space. i.e.,

$$X = H_1(x) = Concat[x, x] \quad (11)$$

where $H_1(x)$ is a concatenation object that consisting of the copied channels.

The proposed HGM-copy involves three characteristics: (i) learning some prior information in higher-dimensional space, rather than the original space; (ii) pre-processing of channel-copy transformation provides richer data samples $n$ without changing the characteristic of the data; (iii) incorporating the higher-dimensional prior into the iterative restoration procedure to handle the original IR problem.

On the other hand, pixel-scale transformation is used as a new representation in HGM-pool to generate latent space obeying Gaussian distribution. Motivated by the popular pooling operator [59], down-sampled operation is used in to produce multi-channel object. Fig. 4 (b) visually illustrates the architecture, i.e.,

$$X = H_2(x) = Concat[x_{p1}, x_{p2}, x_{p3}, x_{p4}] \quad (12)$$

where the down-sampling operator $H_2(x)$ generates four different sub-images, each of them contains 1/4 amount of the pixels in $x$. We take advantage of the multi-scale feature of PST in a "joint" manner by decomposing the image into small blocks information, thus reduces the feasible space dimension from $d$ to $d/4$.

It is predictable that HGM will increase the stability of network training, optimize the convexity of latent space representations, and preserve high-dimensional structures in the latent space. The details of HGM-copy and HGM-pool are tabulated in Table I.

TABLE I
TWO TYPES OF INVERTIBLE TRANSFORMATION IN HGM.

| Description | Function | Reverse |
|---|---|---|
| HGM-copy | $X = H_1(x) = Concat[x, x]$ | $x = H_1^{-1}(X)$ |
| HGM-pool | $X = H_2(x) = Concat[x_{p1}, x_{p2}, x_{p3}, x_{p4}]$ | $x = H_2^{-1}(X)$ |

Originally, NCSN employs 3-channel color image $x = [x_R, x_G, x_B]$ as the input of network to learn the gradients of the data distribution $\nabla_x \log p(x)$. Instead of using an explicit representation of image as the input, HGM will learn the more sophisticated image prior $\nabla_X \log p(X)$.

After the object $X$ is chosen, the high-dimensional distribution is estimated with DSM at the prior learning stage. Specifically, for every continuously differentiable probability density $p(X)$ of $X$, we call $\nabla_X \log p(X)$ as its score function. We implement high-dimensional tensor $X$ as input of the network to learn the gradients of the data distribution $\nabla_X \log p_{\sigma_i}(X)$ by training a single neural network $s_\theta(X, \sigma_i)$ with the following loss:

$$\frac{1}{2L} \sum_{i=1}^{L} E_{p_{data}(X)} E_{p_{\sigma_i}(\tilde{X}|X)} \lambda(\sigma_i) \| s_\theta(\tilde{X}, \sigma_i) + (\tilde{X} - X)/\sigma_i^2 \|_2^2 \quad (13)$$

The essence of assisting to be $X$ is to obtain much more data information at high-dimensional manifold and high-density regions. Thus, it avoids some difficulties in score estimation of DSM [43], [47]. Subsequently, training HGM over data $X$ in high-dimensional space as network input and the parameterized $s_\theta(X, \sigma_i)$ is obtained. The visualization of the prior learning stage is demonstrated in the top region of Fig. 5. In general, the high-dimensional prior information stacked in the channel is used as the input of the network to assist generative the model in training stage. In other words, learning existing information based on higher dimensionality.

HGM fully discovers the relationship between the representation error and sample number, internal dimension, respectively. To be more specific, HGM improves the accuracy of score matching by increasing the sample size and reducing the spatial dimension. Due to the elegant solution regarding issues related to low data density regions and manifold hypothesis, it becomes less sensitive under the influence of saddle points.

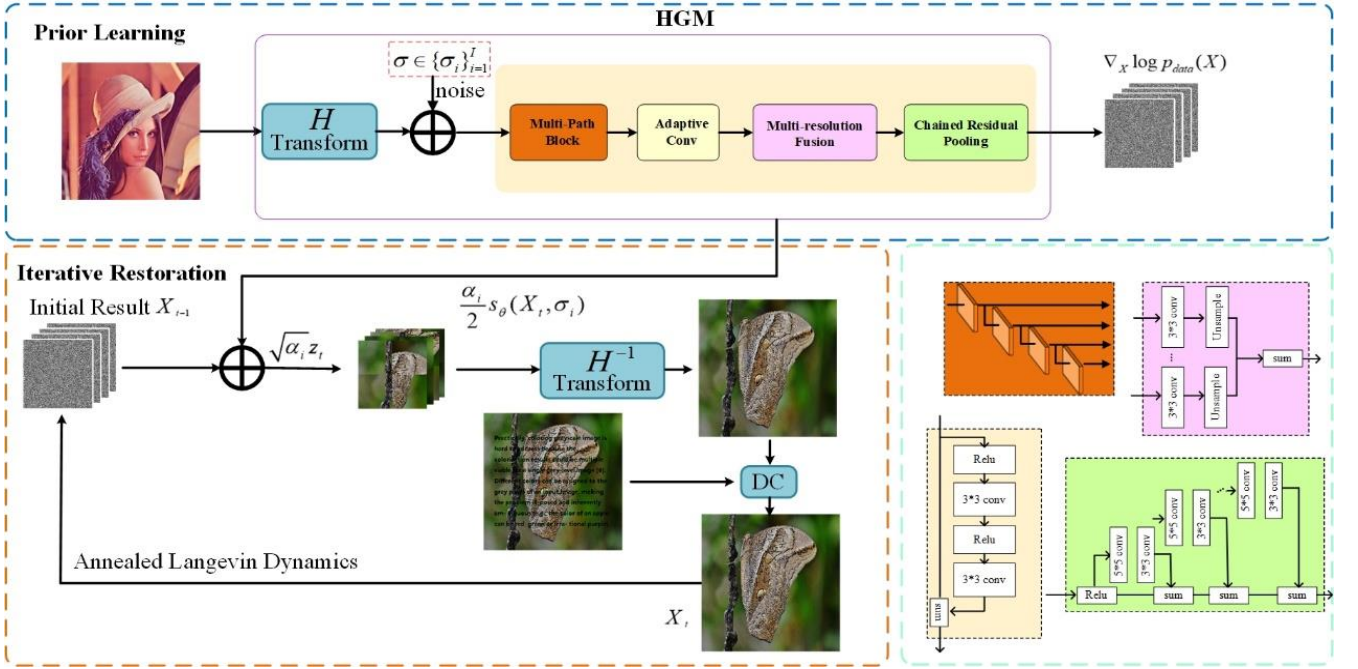

**Fig. 5.** The pipeline of HGM. The blue box is the training process for prior learning, the yellow box is the procedure for IR, and the green box is the detailed structure diagram of HGM. In the training pass, we learn the score of perturbed low-resolution high-dimensional images via $H$ transformation and add a series of noises $z$. At the stage of restoration, the Gaussian noise is gradually approximated to the real score and is sampled by annealed Langevin dynamics. The DC is the LaGrange term to enforce the mixture constraint.

### C. HGM: Iterative Restoration

The entire IR procedure mainly involves two ingredients in the testing phase: Annealed Langevin dynamics and high-dimensional data-fidelity. In this work, we consider two cases: The basic way and the progressive way. Particularly, the progressive way is an advanced version of the basic way and largely depends on the former. The data-fidelity in the progressive way contains the original measurement and the intermediate solution.

**The Basic Way:** Annealed Langevin dynamics is introduced as a sampling approach. Once the score function is known, the Langevin dynamics can be applied to sample from the corresponding distribution. In the iterative restoration process, samples of each disturbance noise level are used as the initial input of the next noise level until reaching the smallest level which enlarges the samples for the network to generate the final restoration result gradually.

Specifically, given step size $\alpha > 0$, the total number of iterations $T$, and an initial sample $X_t$ from any prior distribution, Langevin dynamics iteratively evaluate the followings:

$$X_{t+1} = X_t + \frac{\alpha}{2}\nabla_X \log p(X_t) + \sqrt{\alpha}z_t \quad (14)$$

where $\forall t: z_t \sim N(0, I)$.

Supposing we have a neural network $s_\theta(X)$ parameterized by $\theta$, and it has been trained such that $s_\theta(X, \sigma_i) \approx \nabla_X \log p(X, \sigma_i)$ for all $i \in \{1, 2, \cdots, L\}$. By decreasing the $\alpha_i$-value (accordingly $\sigma_i$-value), we can approximately generate samples from $p(X)$ using annealed Langevin dynamics by replacing $\nabla_X \log p_{\sigma_i}(X_t)$ with $s_\theta(X_t, \sigma_i)$ iteratively, i.e.,

$$X_{t+1} = X_t + \frac{\alpha_i}{2}s_\theta(X_t, \sigma_i) + \sqrt{\alpha_i}z_t \quad (15)$$

Specifically, at each iteration of the annealed Langevin dynamics, we update the solution via data likelihood constraint after Eq. (15). The solving of $x$ can be decomposed to be the alternative minimization procedure between data-fidelity with $x$ and prior information updating with $X$. Typically, the data-fidelity problem is formulated as follows:

$$x_{t+1} = \arg\min_x \lambda \|y - Mx\|^2 + \|x - H^{-1}(X_t)\|^2 \quad (16)$$

The essence of Eq. (16) is to integrate the prior information learned from higher-dimensional space into the lower-dimensional IR problem. The data-fidelity term in Eq. (16) can be solved via the following scheme:

$$x_{t+1} = (\lambda M + I)^{-1}[\lambda M^T y + H^{-1}(X_t)] \quad (17)$$

where $M$ is a binary diagonal matrix. It is an idempotent matrix and $M^T M = M$. The visualization of the iterative restoration stage is demonstrated in the bottom region of Fig. 5.

**The Progressive Way:** To leverage the feasibility of tackling the non-convex minimization, we can turn to the progressive way (HGM-p), i.e., before updating the prior information and data consistency in high-dimensional space, we first pursue the intermediate solution in original space.

For restoration purpose, we estimate the gradients of the RGB data distribution with DSM and then iteratively update samples via Langevin dynamics using those gradients and the data consistency. Specifically, in original space, we use 3-channel color image $x$ as the input of network to learn the gradients of the data distribution $\nabla_x \log p_{\sigma_i}(x)$ and sample $x_{rec}$ by Eq. (6). Annealed Langevin dynamics combined with data-consistency (DC) flow is introduced to generate results that close to the original color gradually. Eq. (15) can be rewritten as follows:

$$X_t \leftarrow X_{t-1} + \frac{\alpha}{2}\nabla_X[\log p(X_{t-1}) - \lambda DC(X_{t-1})] + \sqrt{\alpha_i}z_{t-1} \quad (18)$$

where $DC(X_{t-1}) = H[H^{-1}(X_{t-1}) - y]$. The result is defined as $x_{rec} = y$.

The iterative formula of annealed Langevin dynamics in HGM-p can be rewritten as:

$$X_t \leftarrow X_{t-1} + \frac{\alpha_i}{2}[s_\theta(X_{t-1}, \sigma_i) - \lambda\nabla_X DC(X_{t-1})] + \sqrt{\alpha_i}z_{t-1} \quad (19)$$

Notably, we allow the iterative process to dynamically fine-tune the distribution distance in a progressive regularization manner obtained by HGM. Compared with the non-progressive counterpart, the progressive strategy can better maintain the consistency between the missing distribution and the existing distribution.

---

**Algorithm 1 HGM for IR**

**Initialize:** $\sigma, \sigma^* \in \{\sigma_i\}_{i=1}^L, \varepsilon, L$, and $x_0$
**For** $i \leftarrow 1$ to $L$ **do**
  $\alpha_i = \varepsilon \cdot \sigma_i^2 / \sigma_L^2$, $X_0 = H(x_0)$
  **If in basic way:**
    **For** $t \leftarrow 1$ to $T$ **do**
      Update $X_t$ via Eq. (15) and Eq. (17)
    **End for**
  **If in progressive way:**
    **For** $t \leftarrow 1$ to $T$ **do**
      Update $x_t$ via Eq. (6) and data-fidelity
    **End for**
    **For** $t \leftarrow 1$ to $T$ **do**
      Update $X_t$ via Eq. (19) and Eq. (17)
    **End for**
  $x_t = H^{-1}(X)$
**End for**

---

In summary, as explained in **Algorithm 1**, the whole IR procedure consists of two-level loops: The outer loop handles $\nabla_X \log p_\sigma(X)$ to approximate $\nabla_X \log p_{data}(X)$ of the intensity domains with decreasing $\sigma_i$-value, while the inner loop decouples to be an alternating process of updating estimated gradient of data prior $\nabla_X \log p_\sigma(X)$ and the least square scheme.

For the convenience of notation, the progressive strategy that conducted in HGM-copy and HGM-pool are coined as HGM-copy-p and HGM-pool-p, respectively.

## IV. EXPERIMENTS

### A. Experiment Setup

In this section, a series of experiments are presented to quantitatively and qualitatively evaluate the performance of HGM. Two IR tasks are implemented: Image demosaicking and image inpainting.

1) **Datasets**: All the reported tests are conducted on two datasets. i.e., Microsoft's Demosaicking dataset MSR is used in image demosaicking, and LSUN dataset is used in image inpainting. More specifically, the MSR dataset consists of 200 training images including linearized 16-bit mosaicked input images and corresponding linRGB ground truths. We also enlarge the number of images by horizontal and vertical flipping. The LSUN dataset is a large color image dataset, which contains around one million labeled images for each of 10 scene categories and 20 object categories. Among them, we choose the indoor scene LSUN-bedroom dataset in the experiment, which has enough samples (more than 3 million) and various colors to verify the effectiveness of HGM.

2) **Model Training**: All images are rescaled so that the pixel values are in the interval of $[0,1]$. We choose the number of standard deviations to be $L=10$ such that $\{\sigma_i\}_{i=1}^L$ is a geometric sequence with $\sigma_1 = 1$ and $\sigma_{10} = 0.01$. Notice that Gaussian noise of $\sigma = 0.01$ is al-most indistinguishable to human eyes. When using annealed Langevin dynamics for image generation, parameters are setting as $T=80$, $\varepsilon=2\times10^{-5}$ and uniform noise is treated as our initial samples. For the different variants of HGM, i.e., HGM-copy, HGM-pool, HGM-copy-p and HGM-pool-p, we find that the models are robust w.r.t. the choice of a stable $T$ and $\varepsilon$-value between $5\times10^{-6}$ and $5\times10^{-5}$. Specifically, we train the refinement model using the Adam optimizer with $\beta=0.9$ and an initial learning rate of 0.0001. The training and testing experiments are performed with a customized version of Pytorch on an Intel i7-6900K CPU and a GeForce Titan XP GPU. The model is trained within 200,000 epochs, and the whole training process costs around 4 days on the machine.

3) **Quality Metrics**: In all the reported experiments, two traditional metrics -- Peak Signal to Noise Ratio (PSNR) and Structural Similarity (SSIM) are recorded to evaluate the quality of images restored by different methods. The higher PSNR and SSIM values, the better visual quality with more details. Denoting $u$ and $\hat{u}$ to be the reconstructed image and the ground truth, respectively. The PSNR is defined as:

$$PSNR(u,\hat{u}) = 20\log_{10}[\text{Max}(\hat{u})/\|u-\hat{u}\|_2] \qquad (20)$$

The SSIM is defined as:

$$SSIM(u,\hat{u}) = \frac{(2\mu_u\mu_{\hat{u}}+c_1)(2\sigma_{u\hat{u}}+c_2)}{(\mu_u^2+\mu_{\hat{u}}^2+c_1)(\sigma_u^2+\sigma_{\hat{u}}^2+c_2)} \qquad (21)$$

For the convenience of reproducible research, source code and results of HGM can be downloaded from website: https://github.com/yqx7150/HGM.

### B. Image Demosaicking

In this experiment, the Bayer color filter array is selected as the degrading operator. Subsequently, the image demosaicking of HGM is evaluated on dataset McMaster [50] and Kodak [51], respectively. Additionally, HGM is compared with two state-of-the-art methods: Klatzer [52] and Khashabi [53], and is also intensively compared with the naive NCSN [43].

The average PSNR and SSIM values of results recovered from the above-mentioned methods for two different datasets are recorded in Table II. It can be observed that the PSNR and SSIM values of HGM are outstanding. In short, the SSIM values of HGM are slightly higher than NCSN, and in turns it is better than Khashabi [53] and Klatzer [52]. In terms of PSNR, the average PSNR value of HGM on Kodak reaches 40.65 dB and reaches 38.02 dB on McMaster, which is 0.5 dB higher than NCSN and 3~12 dB higher than both Klatzer and Khashabi. In particular, HGM-copy-p and HGM-pool-p perform better than others by a large margin.

As shown in Figs. 6 and 7, the results by all methods have been significantly recovered, except for Khashabi and Klatzer, which are relatively poor. Observing the details of the local magnifications in Fig. 6, there are many unrestored green spots in both Khashabi and Klatzer, and the depths of color are unequal to the original image. By contrast, the results in original NCSN perform too smooth, indicating that the details are lacking and the colors are lighter than the observation. Obviously, the visual effect of HGM at the last four images is much more natural, which successfully reconstructs the missing colors and has better visual effect. To illustrate the merits of HGM more intuitively, recovered images with richer pattern details are depicted in Fig. 7. As mentioned above, the visual improvement with the HGM is obvious. Since HGM effectively uses high-dimensional tensors as priors, the effectively progressive strategy is certified. The evidence of the above statement can be proven by the outstanding performance of HGM-copy-p and HGM-pool-p. The result of the progressive method can provide more detailed textures, which is more visually pleasing. It is worth noting that the unsatisfied performance of HGM-pool may be due to the fact that the pool operator is not easily addressed in this task.

In summary, there are obvious advantages of HGM from the comparison of quantitative values and vision inspection. Combining the powerful representation ability of the generative model and the rich high-dimensional features in demosaicking task by HGM is valid. In particular, the results of HGM-copy-p and HGM-pool-p are more promising.

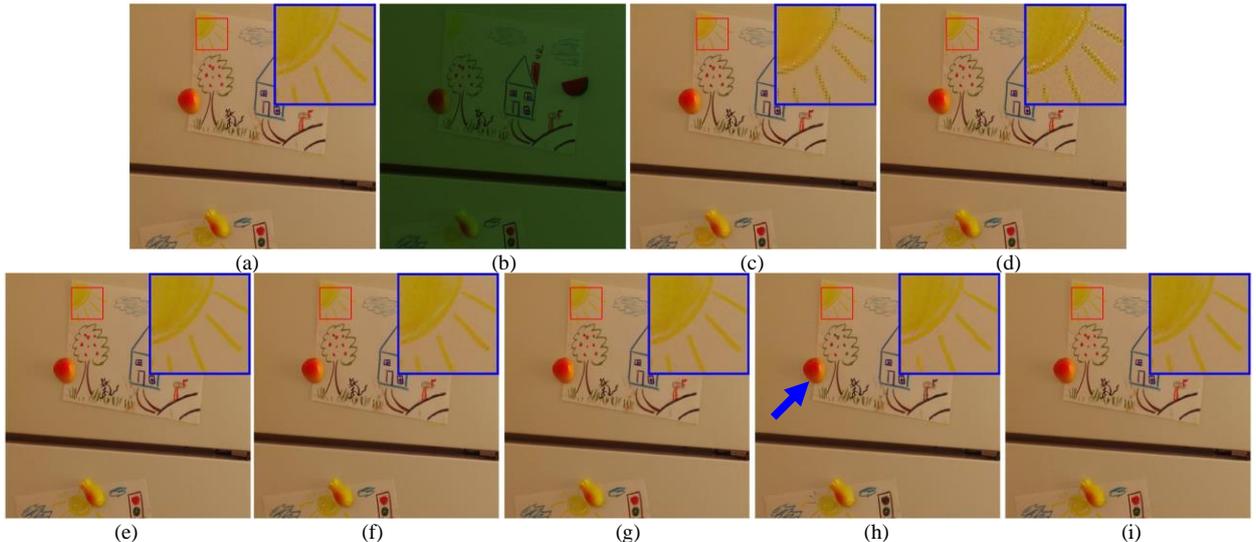

**Fig. 6.** Visualization for demosaicking comparison of HGM with other methods on image from the McMaster dataset. (a) Ground truth, (b) Observation, (c) Khashabi (PSNR = 36.13 dB; SSIM = 0.9339), (d) Klatzer (PSNR = 37.35 dB; SSIM = 0.9449), (e) NCSN (PSNR = 42.26 dB; SSIM = 0.9614), (f) HGM-copy (PSNR = 40.91 dB; SSIM= 0.9577), (g) HGM-copy-p (PSNR = 42.40 dB; SSIM= 0.9623), (h) HGM-pool (PSNR = 34.77 dB; SSIM= 0.9388), (i) HGM-pool-p (PSNR = 43.86 dB; SSIM = 0.9716).

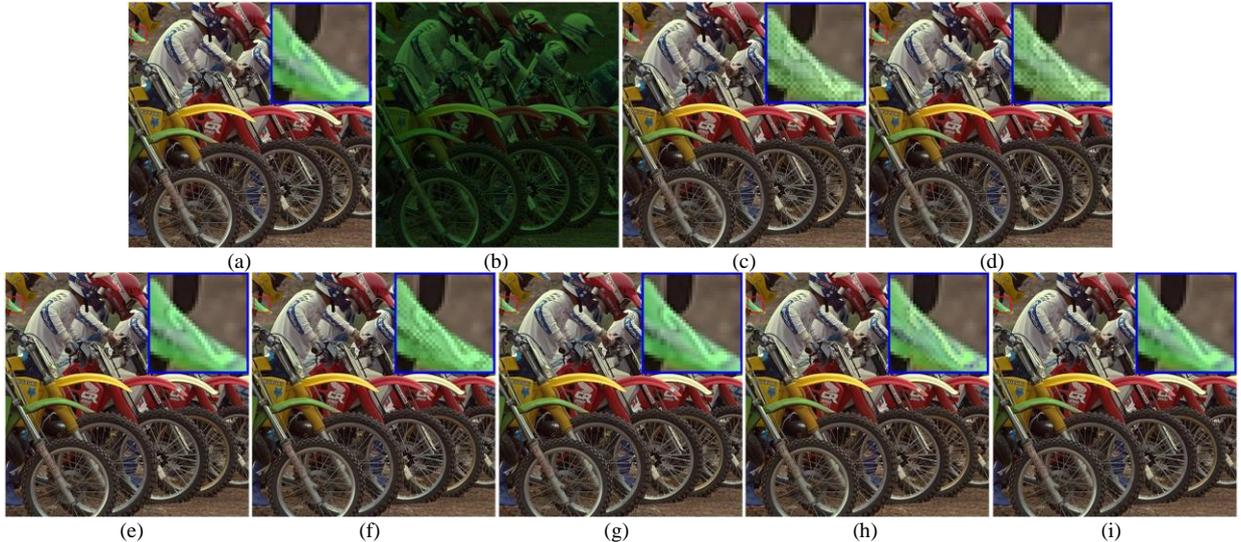

**Fig. 7.** Demosaicking comparison on image from the Kodak dataset. (a) Ground truth, (b) Observation, (c) Khashabi (PSNR = 37.47 dB; SSIM = 0.9748), (d) Klatzer (PSNR = 39.28 dB; SSIM = 0.9874), (e) NCSN (PSNR = 39.46 dB; SSIM = 0.9933), (f) HGM-copy (PSNR = 36.80 dB; SSIM = 0.9883), (g) HGM-copy-p (PSNR = 39.40 dB; SSIM = 0.9933), (h) HGM-pool (PSNR = 32.96 dB; SSIM = 0.9719), (i) HGM-pool-p (PSNR = 42.05 dB; SSIM = 0.9959).

TABLE II
COMPARISON OF HGM TO STATE-OF-THE-ARTS ON DEMOSAICKING IN TERMS OF AVERAGE PSNR AND SSIM PERFORMANCE.

| MSR Dataset | Kodak | McMaster |
|---|---|---|
| Khashabi [53] | 37.5/0.9686 | 25.6/0.9327 |
| Klatzer [52] | 38.9/0.9823 | 35.1/0.9441 |
| NCSN [43] | 40.1/0.9868 | 38.6/0.9728 |
| HGM-copy | 39.0/0.9825 | 36.7/0.9624 |
| HGM-copy-p | 40.5/0.9865 | 38.8/0.9731 |
| HGM-pool | 35.4/0.9642 | 33.2/0.9389 |
| HGM-pool-p | **42.8/0.9897** | **40.9/0.9804** |

### C. Image Inpainting

In this section, HGM is used for the PNG-coded inpainting tasks on 100 selected images from LSUN dataset. In general, the observation can be modeled as the corrupted one by the quantization mask (i.e., $y = M \otimes x$). Specifically, four masks are selected as degrading operators: Block, Text, Random, and Mnist. To verify that HGM can learn generalizable and semantically meaningful image representations, HGM is compared with Liu *et al.* [54], RFR [55], and NCSN in terms of recovery effect.

Average PSNR and SSIM values of each algorithm are listed in Table III. First, in the comparison of filling contents in the Block, Text, and Mnist masks, HGM works better than Liu *et al.* [54] and RFR [55]. Especially in the case of Text2, the PSNR value of HGM-copy-p is 8.7 dB higher than that in Liu *et al.* [54], and 17.8 dB higher than RFR. Second, under the Random mask, Liu *et al.* [54] has higher PSNR values and HGM has higher SSIM values. In addition, HGM-copy, HGM-pool, HGM-copy-p, and HGM-pool-p are superior to the original NCSN.

Previous experiments have demonstrated that the progressive way with high-dimensional prior information is an improvement on image demosaicking. It can better reflect the information of significant changes such as the edge texture of the image. Although the purposes of Image inpainting and image demosaicking have similar characteristics, the difficulties involved in the recovery procedure are different. i.e., the latter is to restore the information of the two channels, and the former is to restore the information of the whole missing region. The above phenomenon implies that the overall effects of HGM-copy-p and HGM-pool-p are better than HGM-copy and HGM-pool.

Although the very recent work Liu *et al.* [54] has obtained competitive or better performance than HGM in certain scenarios such as Random mask, it is an end-to-end supervised learning and needs to retrain the model for special mask. Contrast to the scheme of learning special model for special mask in Liu *et al.* [54], we only need a single model for all the masks.

TABLE III
AVERAGE PSNR AND SSIM RESULTS OF SEVEN METHODS UNDER DIFFERENT MASKS. HERE, FOUR MASKS INCLUDING THE BLOCK, TEXT, RANDOM AND MNIST MASKS ARE INCLUDED.

| Inpainting Mask | | Liu *et al.* [54] | RFR [55] | NCSN [43] | HGM-copy | HGM-copy-p | HGM-pool | HGM-pool-p |
|---|---|---|---|---|---|---|---|---|
| Block | | 36.48/0.9521 | 23.80/0.5887 | 38.12/0.9829 | 39.49/0.9889 | **39.54/0.9890** | 38.23/0.9849 | 38.46/0.9848 |
| Text | Text1 | 43.99/0.9924 | 36.03/0.9739 | 47.83/0.9978 | **48.65/0.9984** | 48.62/**0.9984** | 48.19/09982 | 48.22/0.9980 |
| | Text2 | 38.50/0.9752 | 29.32/0.8906 | 46.76/0.9978 | 47.16/0.9981 | **47.23/0.9982** | 46.71/0.9976 | 47.01/0.9978 |
| Random | 10% | **30.01/0.8508** | 22.55/0.5998 | 25.62/0.8109 | 25.90/0.8313 | 25.89/**0.8317** | 25.34/0.8196 | 25.70/0.8139 |
| | 20% | **31.33/0.8854** | 23.62/0.6082 | 28.69/0.8935 | 29.38/**0.9129** | 29.37/**0.9129** | 28.92/0.8995 | 28.84/0.8962 |
| | 30% | **33.36/0.9279** | 24.28/0.6189 | 31.86/0.9382 | 32.68/0.9551 | 32.66/**0.9557** | 32.38/0.9467 | 31.90/0.9340 |
| Mnist | | 37.82/0.9739 | 31.89/0.9388 | 38.14/0.9867 | 38.37/0.9876 | **38.40/0.9877** | 38.22/0.9870 | 38.31/0.9870 |

More persuasively, the qualitative recovery comparisons of different methods are demonstrated. Fig. 8 shows the comparison result of the inpainting experiments under two kinds of random noise 10% and 30%. From a global comparison, it is intuitive to see that Liu *et al.* [54] and RFR still retain different degrees of noise. Although Liu *et al.* [54] performs well under Random mask in terms of PSNR value, the visual effect is far inferior to HGM. It can be seen in the magnified area, there still exist black spots on the green bed-sheet in Fig. 8(c). Instead, the estimated images attained by the original NCSN and HGM are much pleasuring than those in Liu *et al.* [54] and RFR. Compared with the reference image, the disadvantage is that the result of NCSN is too aggregated causing bed sheet texture becomes too smooth. Obviously, the results of HGM favor to denoising and restoring textures, especially in HGM-copy-p. Such as the alphabets in the enlarged area are visible from the HGM results in the second image, in which HGM-copy and HGM-copy-p are sharper and in line with the visual beauty of the person. Therefore, it indicates that the DC item is helpful for the restoration of edges and details in inpainting task.

Meanwhile, experimental results under the degradation mask of Block that covering 50% of the image are shown in Fig. 9. The degraded image still contains rich color information and high-frequency information. In the two comparative experiments of the first line, the results of RFR still remain traces of the block. Although Liu *et al.* [54] in Fig. 9(c) obtains higher PSNR than RFR, there is still a light faint. In contrast, both NCSN and HGM in the second line of Fig. 9 perform more satisfactorily. Observing the magnified area, HGM generates more semantically plausible and visually pleasing results.

To make a deeper evaluation of our method in terms of visual quality, we imitate the content missing of real data under the mask Mnist. Specifically, we randomly select 100 images from the testing dataset and destroy them with random handwriting to obtain the corrupted Mnist observation. Fig. 10 presents the subjective comparison of recovery images. It can be clearly seen that NCSN does not recover intricate semantic information very well, while HGM-copy-p and HGM-pool-p recover more clear image edges and texture details. For more intuitively comparison, Table III records the average PSNR of the seven methods on five Mnist masks. The PSNR value of RFR is only 31.89 dB and the unremoved noise can be seen in Fig. 10(d). In contrast, the PSNR value of Liu *et al.* [54] on the Mnist mask can reach 37.82 dB, but the outline of the handwriting can be seen vaguely. Totally, the average PSNR of HGM is 0.46 dB higher than Liu *et al.* [54] and 6.4 dB higher than RFR. Thus, HGM achieves preeminent performance in a variety of inpainting masks.

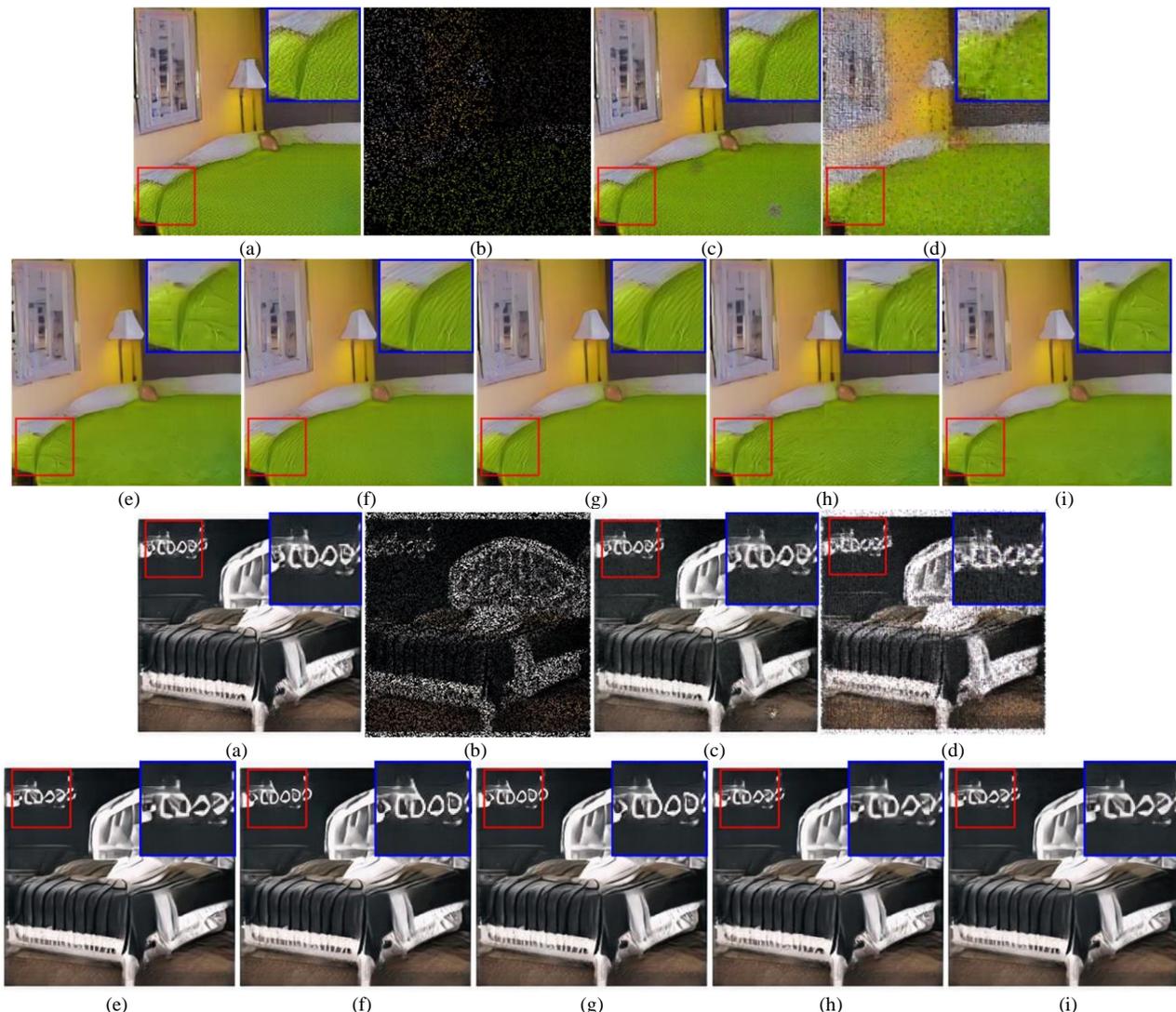

(a) (b) (c) (d)

(e) (f) (g) (h) (i)

(a) (b) (c) (d)

(e) (f) (g) (h) (i)

**Fig. 8.** Subjective comparison for LSUN-bedroom dataset. The results are under the Random mask with 10% and 30% sampling ratios, respectively. (a) Ground truth, (b) Observation, (c) Liu *et al.* [54], (d) RFR, (e) NCSN, (f) HGM-copy, (g) HGM-copy-p, (h) HGM-pool, (i) HGM-pool-p.

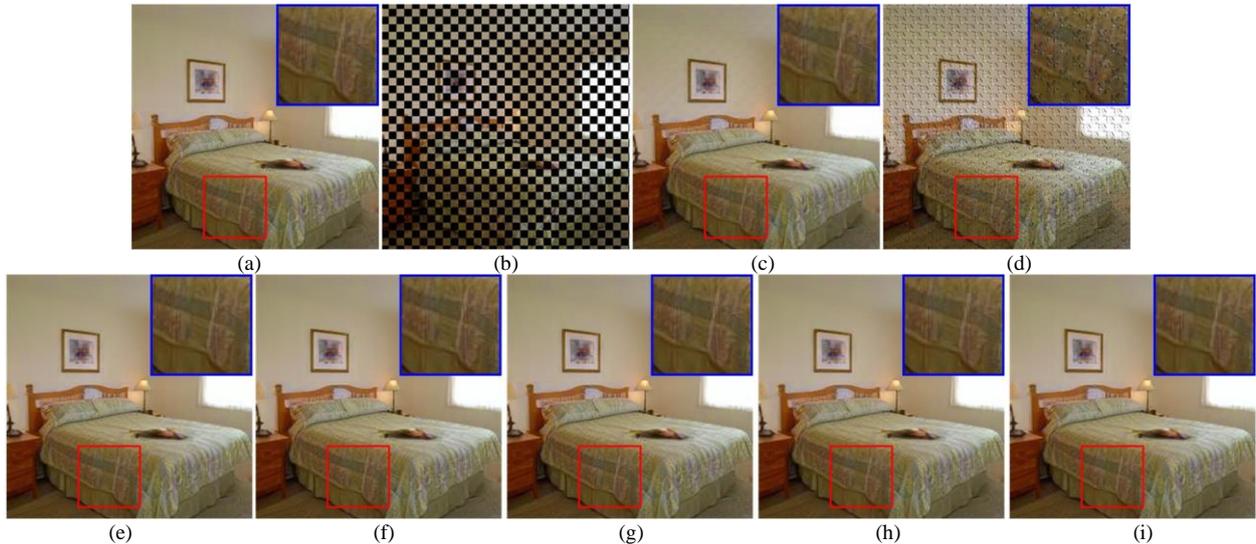

**Fig. 9.** Subjective comparison under the Block mask with 50% coverage. (a) Ground truth, (b) Observation, (c) Liu *et al.* [54], (d) RFR, (e) NCSN, (f) HGM-copy, (g) HGM-copy-p, (h) HGM-pool, (i) HGM-pool-p.

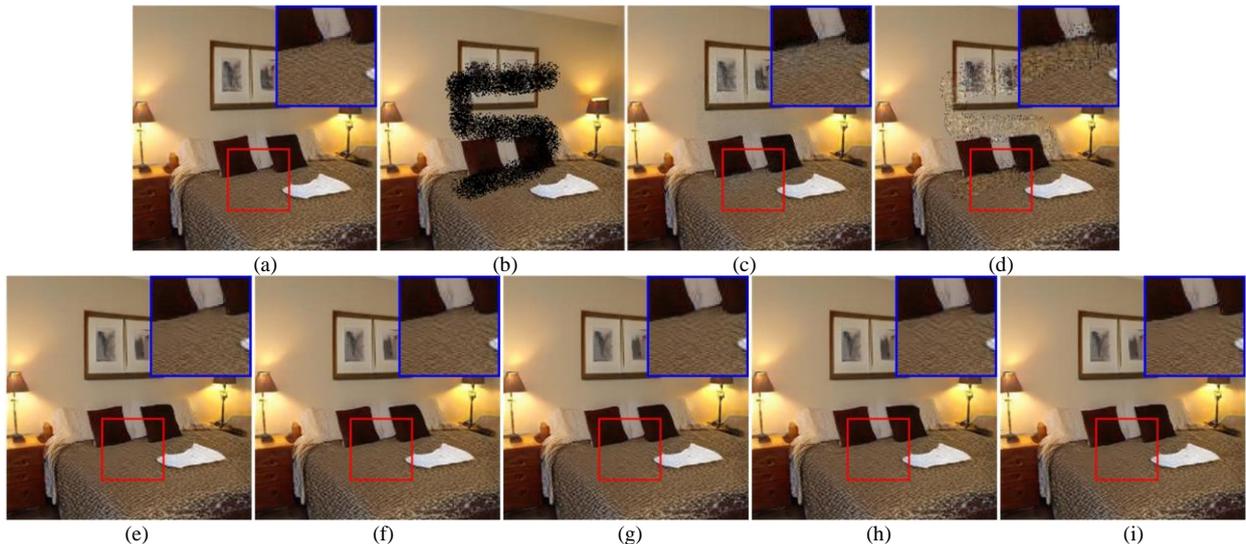

**Fig. 10.** Subjective comparison for LSUN-bedroom dataset under the Mnist mask. (a) Ground truth, (b) Observation, (c) Liu *et al.* [54], (d) RFR, (e) NCSN, (f) HGM-copy, (g) HGM-copy-p, (h) HGM-pool, (i) HGM-pool-p.

*D. Limitations of HGM*

As well known, the larger the mask that needs to be filled, the more serious the ill-posedness will be. In the previous sections, HGM attains promising results in inpainting tasks for filling in relatively small regions. Here, some limitations of HGM will be discussed. More precisely, filling in large blocks will be investigated.

One of the possible advantages in the supervised learning scheme is that it is designed to special task, while lacks flexibility. On the contrary, HGM belongs to the category of unsupervised learning, which obtains prior information through training data, with strong flexibility and wide application scenarios. In the circumstance of the ideal and abundant training data to be available, supervised learning is better than unsupervised learning counterpart. Taking an example for filling large regions, supervised learning will work better. Since the prior information is learned well, HGM can still achieve acceptable and reasonable results.

To demonstrate the above statement more intuitive, the objective results of PSNR and SSIM values are given in Table IV. Among them, the average PSNR value under HGM is 37.15 dB in the Curve degradation mask, while the PSNR of Liu *et al.* [54] is 41.18 dB. Due to the presence of irregular texture patterns, the filling effect under the missing large area is not as good as both the Textural mask and randomly sampled masks and does not satisfy the human visual experience. Fig. 11 provides the subjective comparison under curve mask. The filling effect of HGM is not as brilliant as *Liu et al* which using supervised filling method. Nevertheless, HGM still fills the missing area reasonably and effectively.

TABLE IV
AVERAGE PSNR AND SSIM VALUES OF HGM IN COMPARISON WITH STATE-OF-THE-ART TECHNIQUES ON FILLING LARGE REGION.

| Method | Curve mask | Method | Curve mask |
|---|---|---|---|
| Liu *et al.* [54] | **41.18/0.9908** | HGM-copy | 36.64/0.9857 |
| RFR [55] | 31.35/0.9326 | HGM-copy-p | 36.97/0.9869 |
| NCSN [43] | 37.64/0.9870 | HGM-pool | 36.86/0.9863 |
| ------ | ------ | HGM-pool-p | 38.14/0.9890 |

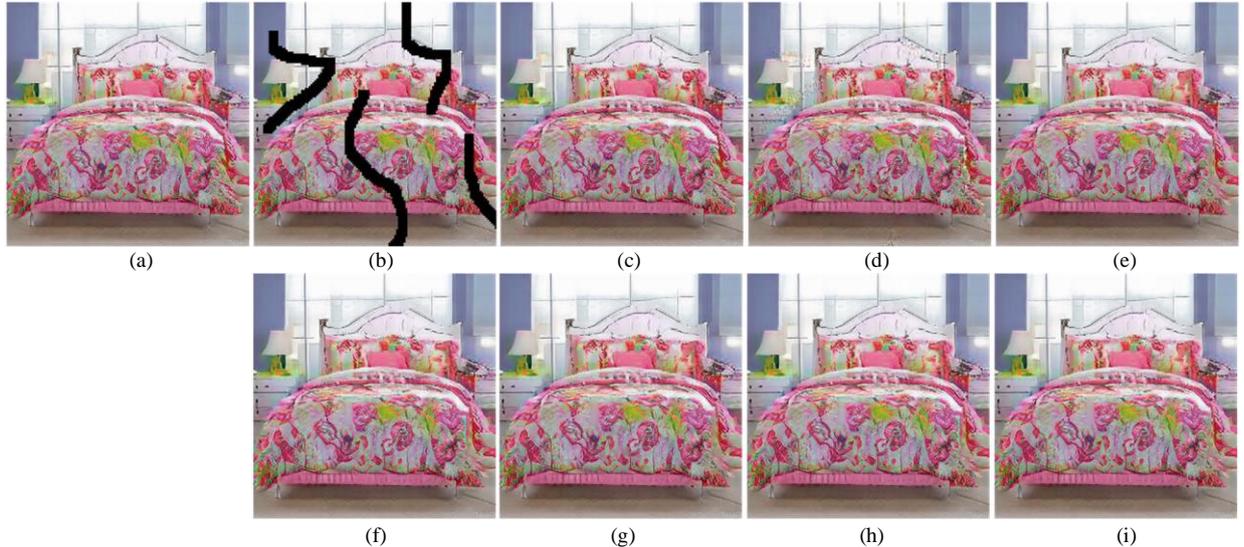

Fig. 11. Subjective comparison under the Curve mask. (a) Ground truth, (b) Observation, (c) Liu *et al.* [54], (d) RFR, (e) NCSN, (f) HGM-copy, (g) HGM-copy-p, (h) HGM-pool, (i) HGM-pool-p.

## V. Discussions

Two ways for constructing high-dimensional signals are discussed in this study: HGM-copy and HGM-pool. It can also be implemented in wavelet domain that makes full use of multi-resolution and multi-scale characteristics. Moreover, irreversible gradient transformation is also a possible alternative.

**Invertible Wavelet Transformation.** Over the past decades, invertible discrete wavelet transformations (DWTs) have rapidly developed from math to engineering. Specifically, wavelet transformation can increase feature channels and down-sample feature maps. After the wavelet transformation, the input image $x$ with the size of $m_1 \times m_2 \times c$ is transformed into a new feature map of size $(m_1/2) \times (m_2/2) \times (4c)$. Wavelet decomposes an input image into one low-frequency representation and three high-frequency representations in the vertical, horizontal, and diagonal directions, i.e., $W(x) = \{LL, LH, HL, HH\}$. Conversely, the original input image can be reconstructed by the inverse DWT using $x = W^{-1}(LL, LH, HL, HH)$. Like the proposed HGM-pool, DWT decomposes a high-dimensional tensor into 4 smaller tensors, reducing the feasibility dimensions $d$ through the decomposition operation.

**Irreversible Gradient Transformation.** In traditional algorithms, using gradients to describe images can retain the contrast information of the image and reduce the amount of complexity. The calculation of the image gradient uses the convolution kernel to convolve the image: $\nabla x = I \times W_x$, $\nabla y = I \times W_y$, where $I$ is the input image, $W_x$ represents the convolution operator in the horizontal direction, and $W_y$ represents the convolution operator in the vertical direction. HGM-copy simply uses the copy operation to increase the sample number $n$, but the latent dimension of the data will not change. Gradient transformation can effectively overcome this shortcoming. Since the gradient map after the gradient transformation has a new data density, thus provides a more advantageous over only increasing the sample number. While the irreversibility of gradient transformation is a deficiency that needs to overcome. Although there are methods that can achieve the pseudo reversible, the information loss still exists.

## VI. Conclusions

Based on the observation that the representation error of the prior knowledge from generative models is bounded by sample number and space dimension, in this study, a high-dimensional assisted tensor as the image prior generation model was proposed. This work exploited two ways to incorporate higher-dimensional prior information into the lower-dimensional recovery procedure: HGM-copy and HGM-pool. More specifically, HGM-copy stacked the channel by copying transformation to increase the sample number, and HGM-pool reduced the feasible space dimension. Score-based generative modeling for IR first estimated the gradient of high-dimensional data density through score matching and then generated samples through annealing Langevin dynamics. Moreover, progressive ideology was applied to enhance the learning ability. Finally, experimental results demonstrated that HGM can restore high-quality images in terms of visual inspection and quantitative measurements.